\documentclass{article}
\author{Rasmus Mackeprang (Niels Bohr Institute) \and Andrea Rizzi (Scuola Normale Superiore)}
\usepackage{epsfig,subfigure,amsmath,amssymb,color,feynmp}
\newcommand{\pp }{\ensuremath{2 \rightarrow 2}{ }}
\newcommand{\ppp }{\ensuremath{2 \rightarrow 3}{ }}

\newcommand{\massG}{\ensuremath{\textrm{GeV}/\textrm{c}^2}}
\newcommand{\massT}{\ensuremath{\textrm{TeV}/\textrm{c}^2}}
\date{\today}

\begin{document}
\title{Interactions of Coloured Heavy Stable Particles in Matter}
\author{Rasmus Mackeprang \\(Niels Bohr Institute, Denmark) \and
Andrea Rizzi \\(INFN Pisa,Italy)}

\maketitle
\abstract{ In this paper we present a physics model for the
  interactions of stable heavy hadrons containing a heavy parton with
  matter. The model presented is a natural continuation of the work
  started in \cite{Kraan:2004tz}. However, changes and generalisations
  have been made allowing for the description of a broader scope of
  physics scenarios.
%
  As a special case the model is tested on the cases of stable gluino
  and stop hadrons, thus covering both stable colour triplet and octet
  states. Conclusions are drawn regarding the phenomenology of these
  cases including how to distinguish between them.
}

\section{Introduction}
\label{sec:Intro}
Many extensions to the Standard Model (SM) feature long-lived
particles carrying SM charges. Most prominently Supersymmetry (SUSY)
models exist where the lightest supersymmetric particle (LSP) carries
either electromagnetic or colour charge. Also more generic models
incorporating hidden valleys and global symmetries exist which allow
for similar phenomenological consequences (\cite{Strassler:2006qa},
\cite{Strassler:2006im}).
The particle contents and possible interactions of different physics
scenarios leads to differences in the energy deposition
characteristics of the heavy hadronic states. In this paper we
demonstrate that these differences can be used as a powerful tool to
discriminate between models.

Section \ref{sec:Models} contains a brief summary of contemporary
models providing long-lived or even stable strongly interacting
particles. The models listed in no way comprise a complete list.

The interactions of stable, heavy and coloured particles differ
dramatically to those of the SM particles. For the future possible
discovery of such particles it is therefore crucial to have an
understanding of their interactions with matter.
In section \ref{sec:IntMod} we present a toolkit for simulation of the
interaction of heavy quasi-stable coloured particles with matter. The
physics treatment follows that of the gluino R-hadron study presented
in \cite{Kraan:2004tz} with the difference being the generalisation to
a broader physics scope and the implementation into Geant4
\cite{Agostinelli:2002hh}. Also the kinematical treatment of
secondaries from nuclear interactions differs between the two models.
The structure of the Geant4 framework enables us to easily study heavy
colour triplet and octet states.

Simple phenomenological conclusions for both types of states are
derived from our simulations and presented in section
\ref{sec:results} while the reader is referred to \cite{doc} for usage
documentation. The main difference between colour triplet and octet
states
turns out to be the relation between the electromagnetic and the
hadronic energy loss and the shape of the resulting
dE/dx-distri\-bu\-tions as well as their absolute normalisations. In
the case of models where the heavy states are produced in pairs more
information may of course be deduced from event variables such as
charge correlations between the two particles.

\section{Models with heavy stable coloured particles}
\label{sec:Models}
Several extensions to the SM predict new stable or quasi-stable
interacting particles with high masses. From a collider experiment
point of view quasi-stable particles, i.e. particles decaying with a
$c\tau$ at the order of a few metres, can be regarded as stable.  It
is straightforward to include the interaction with matter of heavy
stable particles that only have electromagnetic interactions (like
stable sleptons) in a detector simulation. The interaction of coloured
particles, however, is not modelled in the current version of the main high
energy physics simulation toolkit, Geant4.

In the following we will provide some examples of theoretical models
predicting heavy stable (or at least long-lived) coloured particles. A
comprehensive overview of different models may be found in
\cite{Fairbairn:2006gg}. Whereas the interaction model presented in
section \ref{sec:IntMod} does not depend on the specific physics
scenario, this may indeed affect the production cross section and
kinematic distributions.

\subsection{Split Supersymmetry}
\label{sec:ssusy}
Long-lived gluinos are predicted by several supersymmetric models
\cite{Baer:1998pg} in which the gluino is the Lightest Supersymmetric
Particle (LSP) or the Next to LSP (NLSP). It is therefore by R-parity
stable or \emph{quasi}-stable. Recently the Split supersymmetry model
\cite{Giudice:2004tc} has predicted potentially long-lived gluinos
with a mass of order 1 \massT. In this scenario the decay of the
gluino into the LSP together with a quark anti-quark pair is heavily
suppressed by a very high squark mass.
\begin{figure}[!htbp]
  \centering
  \input{gdecay.inc}
\ \\
\ \\
  \caption{\protect\footnotesize The decay of a gluino in Split SUSY. The mass term in the
    squark propagator suppresses the decay. }
  \label{fig:gdecay}
\end{figure}
Gluinos in this mass range may be produced in the near future at the
LHC proton-proton collider and, being coloured, they would hadronise
into heavy charged or neutral bound states. Such hadrons are commonly
referred to as R-hadrons as their stability is rooted in R-parity.
R-hadrons can be of several types: gluon-gluino balls, gluino
R-baryon and gluino R-meson. The interaction of gluino R-hadrons with
matter has already been studied in \cite{Kraan:2004tz} and implemented
in a Geant3 toolkit, so we will focus on R-hadrons in the
demonstration of the implemented Geant4 model to easily compare the
results. Some Monte Carlo generators already exist
\cite{Sjostrand:2006za}, \cite{Corcella:2000bw}, that are able to
simulate the production and subsequent hadronisation of gluinos.
These programs in connection with the toolkit presented here make it
possible to study the full range of observables of R-hadron events as
they might appear in a detector.

\subsection{Other models}
\label{sec:LED}
Another model in which a heavy coloured particle is stable is the
extra dimensional SUSY scenario proposed in \cite{Barbieri:2002uk}. In
this model the supersymmetry breaking is obtained with a bound
condition on a compactified extra dimension and the predicted LSP is,
in a large fraction of the parameter space, the stop. As is the case
for gluinos the stop may be produced at LHC and will hadronise in
baryon and meson like states in which one quark or anti-quark is
replaced with the stop or anti stop. Also non supersymmetric models
exists in which stable massive coloured particles are expected.
Universal Extra Dimension (UED) scenarios may contain stable or
quasi-stable Kaluza Klein excitations \cite{Appelquist:2000nn},
\cite{Agashe:2004bm}.  Specifically if the KK number is not violated
(or weakly violated) the excitations may be produced in pairs and they
can then be stable or long lived.

\section{Interaction Model in Geant4}
\label{sec:IntMod}


\subsection{Geant4 framework}
Geant4 is an Object Oriented physics simulation toolkit that is the
current standard for HEP experiments. The previous version, Geant3,
was FORTRAN based and was extensively used by LEP experiments and in
the preparation of the LHC experiments. Due to Geant4 being an OO
toolkit the interaction model was implemented as a simple add-on. This
meant that no modifications to the core of the toolkit were required.
The implementation included adding new particles (e.g. R-hadrons) and
their corresponding interaction processes. A particle in Geant4 is
represented by a pair of objects, a G4ParticleDefinition and a
G4DynamicParticle.  The G4ParticleDefinition contains the static
information about the particle type such as name, mass, PDG code, spin
etc. The G4DynamicParticle contains the four momentum of a specific
particle in an event. The new package adds particle definitions for
all particles in the relevant physics scenario. The particles are
defined from a file that contains PDG-code of the particle, mass and
name in a Susy Les Houches Accord (SLHA) like format
\cite{Skands:2003cj}.  The particle definitions are all instances of a
class called CustomParticle that not only contains the normal static
information but also allows the splitting of the particle into a heavy
parton (e.g.  a gluino) and a light quark system (LQS). This is used
to facilitate the treatment of interactions with matter in an easy and
correct fashion.  The interactions with matter are described in a so
called G4Process. A dedicated G4Process has been implemented as
described in following sections. In order to let Geant4 know of this
process it should be registered to the Geant4 framework for each
particle that undergoes this process. This must be done when
compiling, but a great amount of flexibility may be retained for
subsequent run-time reconfiguration.

\subsection{Cross Section}
\label{sec:xsec}

Barring one extension, the cross section calculation follows closely
that of \cite{Kraan:2004tz}. We therefore refer to that paper for a
thorough review.

The basic idea is to use a purely geometric cross section tuned to
$\pi-p$ scattering. The heavy parton is neglected in this context. As
the spatial extent of a wave function scales as $1/M^2$, the cross
section contribution becomes negligible. In this picture we are thus
considering the heavy parton to be exclusively a reservoir of kinetic
energy.
In contrast, each light valence quark is assigned a total cross
section of 12 mbarn per nucleon leading, for instance, to a value of
24 mbarn for a meson carrying a colour octet heavy parton (e.g. a
gluino R-meson).  This value stems from the fact that the size of the
total cross section should be matched to the high energy limit of the
total $\pi-p$ cross section. The contribution for an $s$ quark is set
to 6 mbarn to conform with \cite{Kraan:2004tz}.
For the translation of a cross section per nucleon, $\sigma_{n/p}$,
into a cross section per nucleus, $\sigma_A$, the same GHEISHA
convention \cite{gheishamanual} is used as in the Geant3 work. The
relation between the two cross sections is:
\begin{equation}
  \label{eq:xsectrans}
  \sigma_A = 1.25\times\sigma_{n/p}\times A^{0.7}
\end{equation}

\begin{figure}[!htbp]
  \centering
  \subfigure[In Centre of Mass System (CMS)\label{fig:xsec_cms}]{\epsfig{file=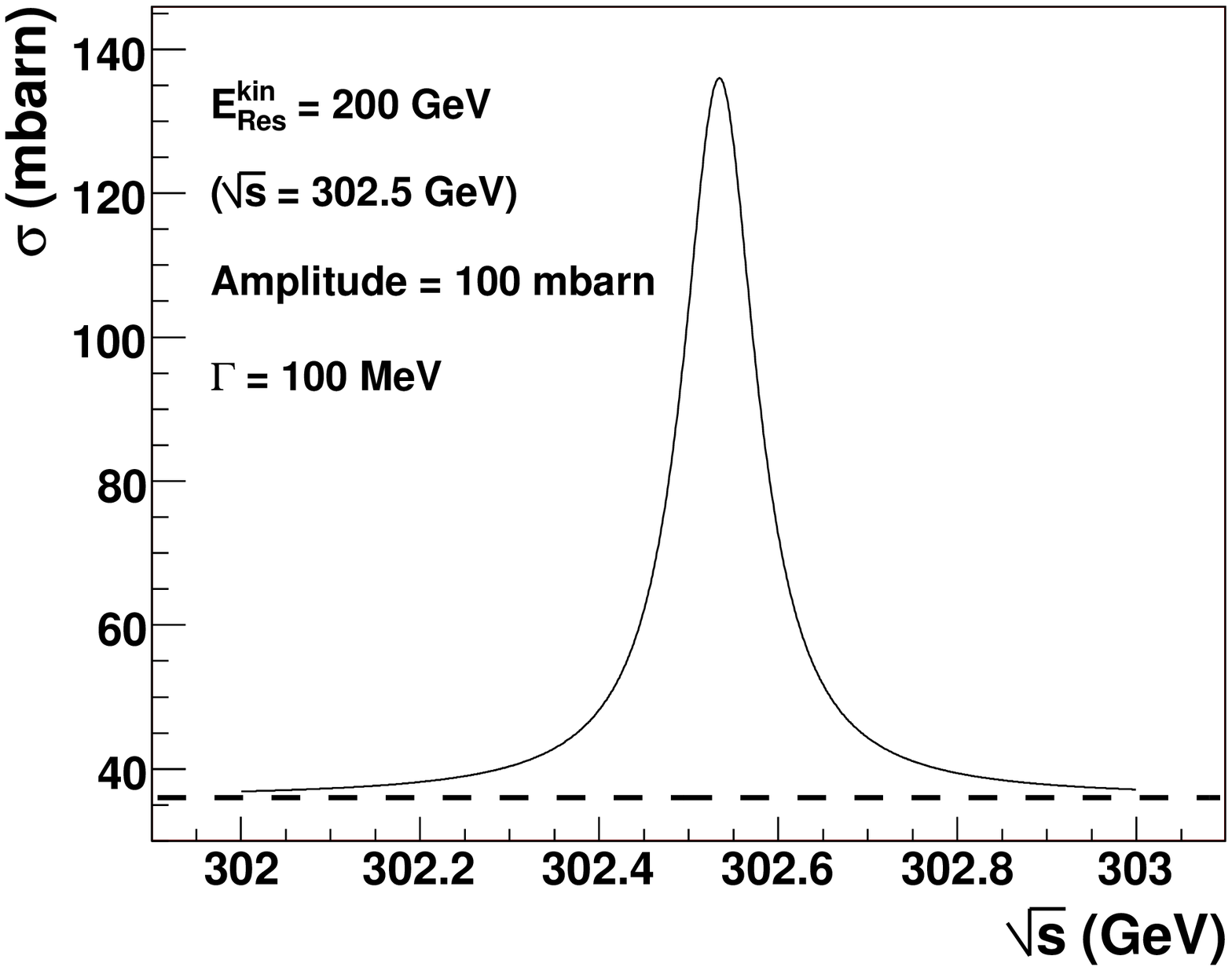,width=6cm}}
  \subfigure[In labframe\label{fig:xsec_lab}]{\epsfig{file=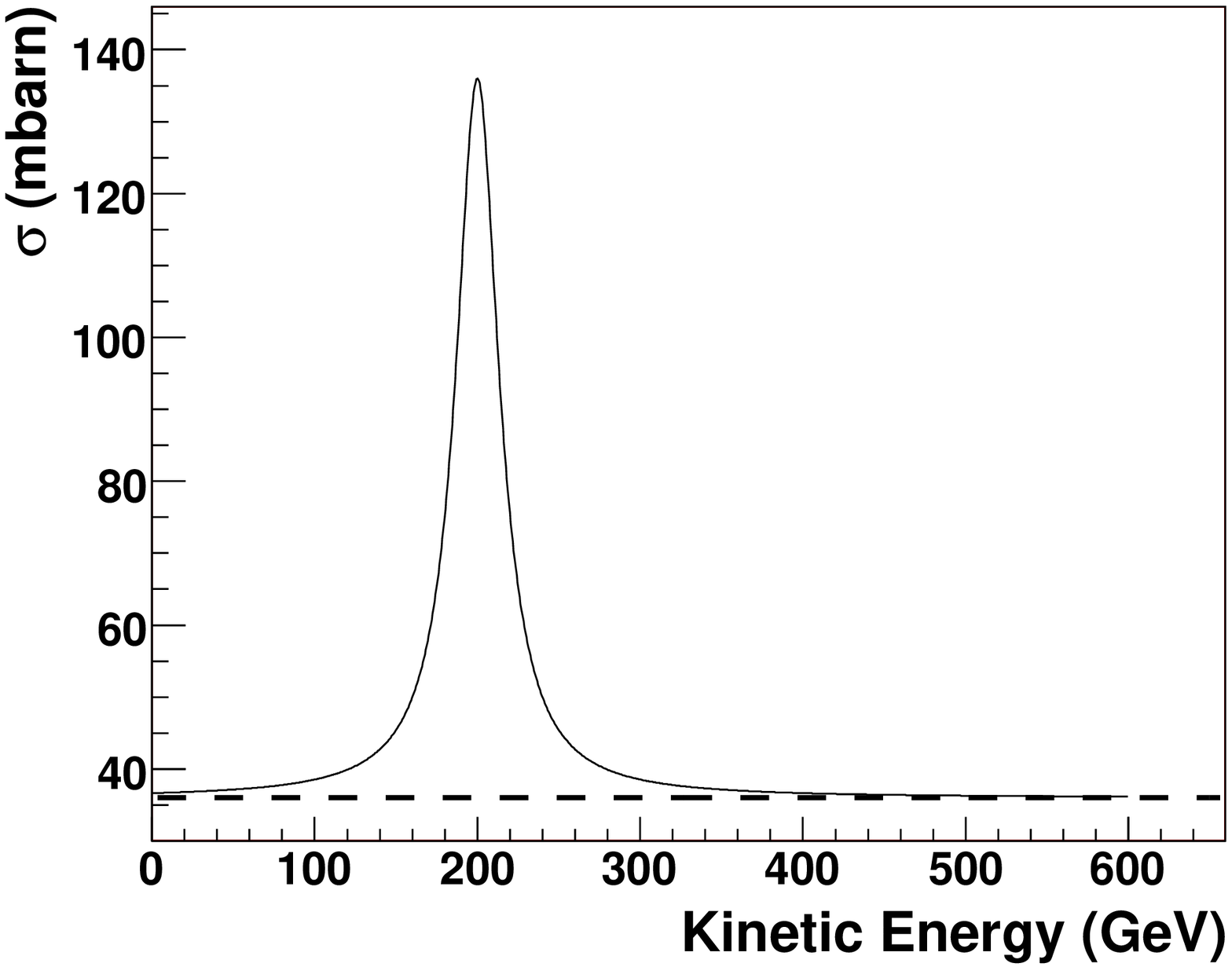,width=6cm}}
  \caption{\protect\footnotesize An example showing how a resonance
    could be configured.  This example is a gluino R-baryon
    ($m_{\tilde{g}}=300\  \massG$) with a base cross section of 36 mbarn
    and a 100 mbarn resonance added.}
  \label{fig:xsecres}
\end{figure}

One additional feature that has been introduced is the ability to add
to the flat cross section a non-relativistic Breit-Wiegner resonance.
The user may at run-time specify resonance position, width and height as
shown in figure \ref{fig:xsecres}. 
\subsection{Choice of Final State}
\label{sec:fstate}

\begin{figure}[htbp]
  \centering
  \epsfig{file=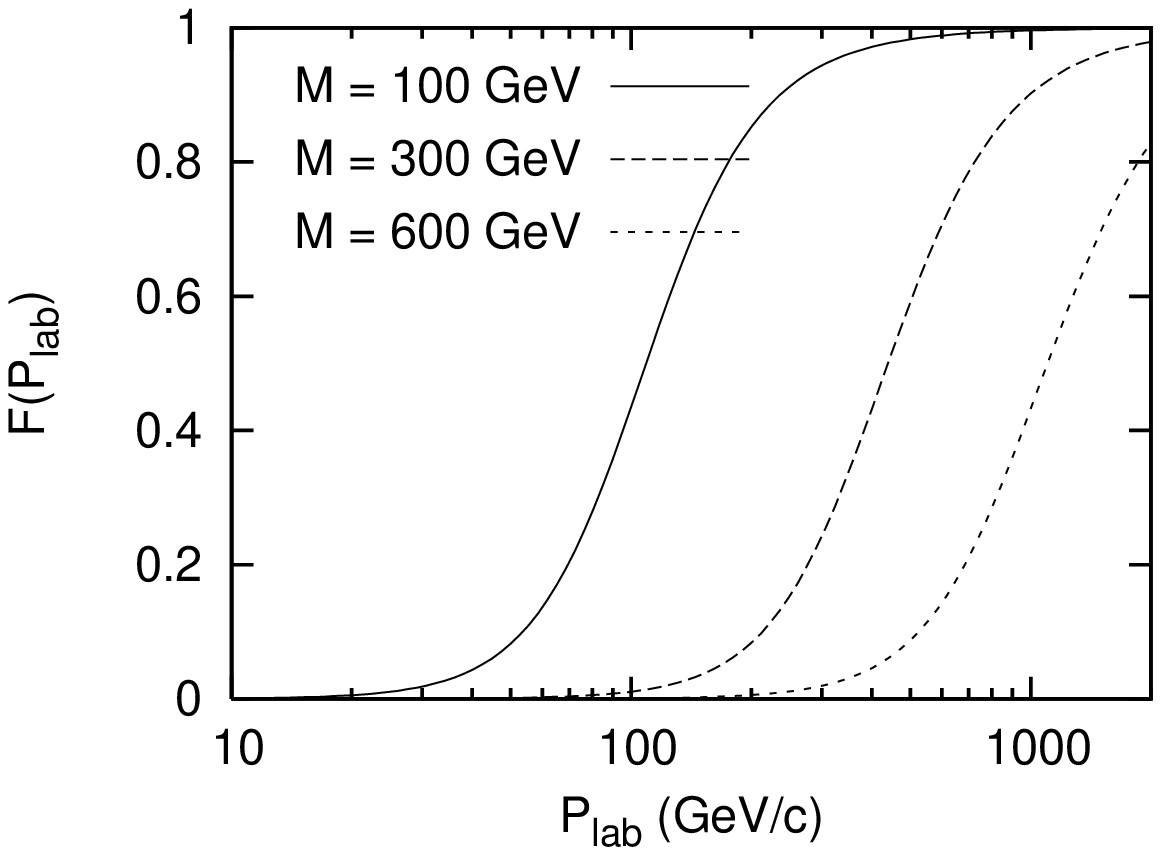,width=4cm,angle=-90}
  \epsfig{file=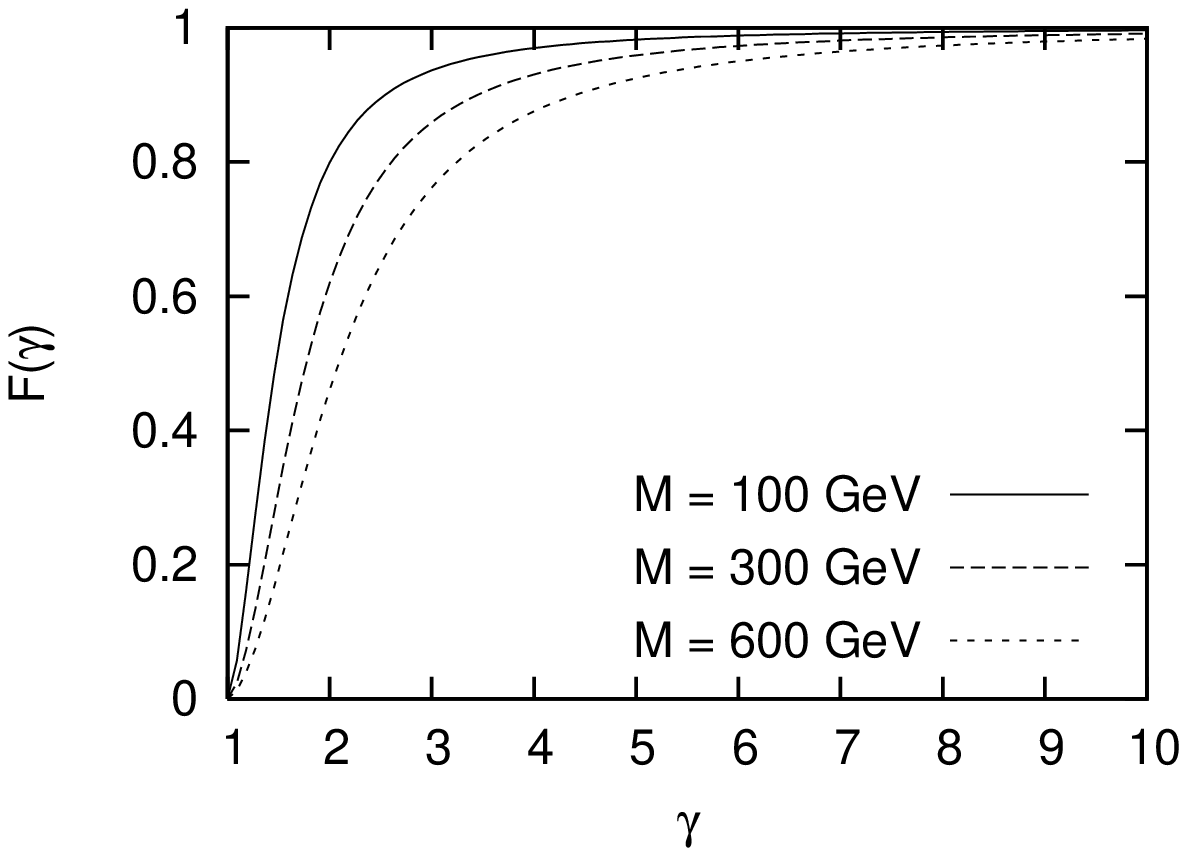,width=4cm,angle=-90}
  \caption{\protect\footnotesize Phase space function $F(Q)$ shown in
    a few parametrisation.
  }
  \label{fig:phase}
\end{figure}

The choice of final state is made using a slightly modified decision
tree with respect to the one used in the Geant3 work, although the end result is
the same. The decision tree is an iteration over the following steps
until a process is selected.
\begin{itemize}
\item Populate a list of possible processes (positive Q-value) \cite{doc}
\item Assign \pp and \ppp processes a priori relative probabilities of
  15\% and 85\% respectively, as elastic scattering amounts to
  approximately 15\% of the total cross section for light hadrons at
  high momenta.
\item Choose a process at random, given the a priori probabilities
\item For \ppp processes, a phase space function decides the
  probability of the process being accepted.
\end{itemize}
The phase space function describes the phase space fraction available for
$2\rightarrow 3$ scattering. It is defined using the same expression as in Geant3:
\begin{equation}
  \label{eq:phase}
  F(Q) = \frac{\sqrt{1+\frac{Q}{2m_{\pi}}}\left(\frac{Q}{Q_0}\right)^{3/2}}{1+\sqrt{1+\frac{Q}{2m_{\pi}}}\left(\frac{Q}{Q_0}\right)^{3/2}}
\end{equation}
This choice of functional expression is motivated by the fact that the
function is, by Lorentz invariance, expected to be a function of
$Q=\sqrt{s}-\sum_i m_i$ where $i$ enumerates the final state
particles. As an ansatz one may use the expression
\begin{equation}
  \label{eq:absphase}
  F(Q) = \frac{\frac{\textrm{d}\phi_3(Q)}{\textrm{d}\phi_3(Q_0)}}{\frac{\textrm{d}\phi_2(Q)}{\textrm{d}\phi_2(Q_0)}+\frac{\textrm{d}\phi_3(Q)}{\textrm{d}\phi_3(Q_0)}}
\end{equation}
where $\textrm{d}\phi_N(Q)$ denotes the \emph{absolute} phase space
for N-body scattering at a scale $Q$. Note that the phase spaces are
normalised to a scale $Q_0$ due to the fact that they have different
dimension. The actual derivation of equation (\ref{eq:phase}) is
rather involved \cite{Kraan:thesis} and involves some
approximations. The value of $Q_0$ is determined by imposing that
$F(Q_0)=0.5$ as required in equation (\ref{eq:absphase}). It turns out
that the value becomes 1.1 GeV. 
F is shown as a function of some
commonly used parameters in figure \ref{fig:phase}.

\begin{figure}[!htbp]
  \centering
  \epsfig{file=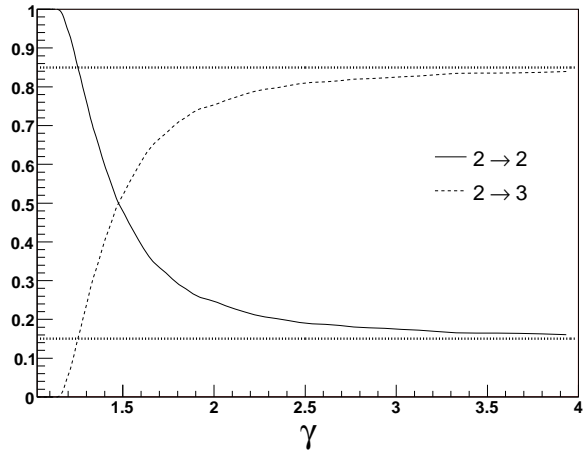,width=8cm}
  \caption{\protect\footnotesize Fraction of $2\rightarrow 2$ and $2\rightarrow 3$ processes
    selected as a function of $\gamma$ of the incident particle. There
    is no discernible difference between different particle masses or
    types. Although this specific plot was made for 300 \massG{} gluino
    baryons in iron, stop hadrons at other masses in other materials
    yield the same result. }
  \label{fig:multiratio}
\end{figure}

The relative frequencies with which \pp and \ppp processes are chosen
may be seen in figure \ref{fig:multiratio}.

\subsubsection{Charge Exchange Suppression}
In a nuclear interaction a hadron containing a heavy parton may
interact through pomeron exchange or reggeon exchange. Pomeron
exchange will leave the charge unaltered whereas Reggeon exchanges
will allow the charge to be transferred.

In the above discussion no preference has been selected between
reggeon and pomeron exchange. This is of course due to the fact that
we have no a priori reason to assume a difference in strength between
the two mechanisms. Theoretical arguments might appear, however, as to
why one exchange mechanism should dominate over the other and so an
additional feature has been added to supply a charge exchange
suppression mechanism.

The mechanism is simply expressed as a probability of rejecting a
process and rerun the final state selection if the selected final
state changes the charge of the heavy hadron. Other ways to do this
could be devised if theoretical basis could be found for a specific
mechanism.

\subsection{Kinematics}
\label{sec:kin}

Once the final state of an interaction is selected, several things may
occur. Decisions need to be taken as to how momenta are assigned to
the outgoing particles. Also the struck nucleus may undergo some
changes. These decisions are highly model dependent and two models are
presented here.

The basic tenet of both models is that the interaction between the
heavy hadron and the nucleon takes place via the LQS alone. This is a
logical extension of the geometric cross section calculation in
section \ref{sec:xsec}. As the heavy parton in this formulation is
strictly a reservoir of kinetic energy, it has been assumed that it
will not interact. Rather the LQS is assumed to interact exclusively.
Imposing the requirement that the LQS and the heavy parton are
co-moving leads to the condition that the LQS carries the kinetic
energy $E_{kin,LQS} = \frac{M_{LQS}}{M_{tot}}$, leading to the
treatment as a low-energy collision in spite of the high energy
carried by the hadron as a whole.

This treatment is naturally only valid as long as $M_{LQS} \ll M_{Tot}$.

\subsubsection{Simple Toy Model}
\label{sec:toymodel}

This model embodied in the ``ToyModelHadronicProcess'' is included in
the toolkit for scenarios where the user wishes to make initial
conclusions while minimising the ``black box'' of the machinery and
thus maintaining an overview. It is hardly valid for physics
conclusions but it is a useful tool to test the process lists supplied
as it is the simplest ansatz possible for a kinematics treatment.

The collision is treated as that between a free nucleon and the LQS of
the heavy hadron. Quasi-elastic (\pp) processes are handled by the
boosting of the LQS and the target nucleon to the CMS and assigning
the outgoing particles back to back momenta along a random axis.
Momenta are then boosted back to the lab-frame.  Upon creation of
secondaries and assignment of momenta, the heavy parton and the LQS
are re-coupled whilst imposing three momentum conservation and
calculating the energy by keeping the heavy hadron on-shell. This
re-coupling procedure results in an excess energy. The difference may
be quantified by:
\begin{equation}
  \label{eq:virt}
  \Delta E = E_{LQS} + E_{HP} - \sqrt{(\vec{p}_{LQS} +
    \vec{p}_{LQS})^2 + M_{Tot}^2}
\end{equation}
This excess energy is added as a local energy deposit in Geant4. \ppp
processes are treated by generating the momenta as a flat distribution
in a Dalitz plot. The same recombination procedure applies as for the
\pp processes.

\subsubsection{Parametrised Model}
\label{sec:parmmodel}

This model - implemented in the FullModelHadronicProcess - is based
upon the parametrised model for light hadrons \cite{G4Phys} with
certain revisions made to the G4ReactionDynamics class to eliminate
flavour dependence \cite{DW}.  As in the toy model, the heavy parton
and the LQS are decoupled from one another, sharing the available
kinetic energy. The quark system is then passed on to the parametrised
model to generate the secondaries as well as the ``black track
particles''.  These black track particles are nuclear fragments from
the struck nucleus such as deuterons, tritons or $\alpha$ particles
created with very low energies.

\begin{figure}[!htbp]
  \centering
  \subfigure[Gluino hadrons in
  iron]{\epsfig{file=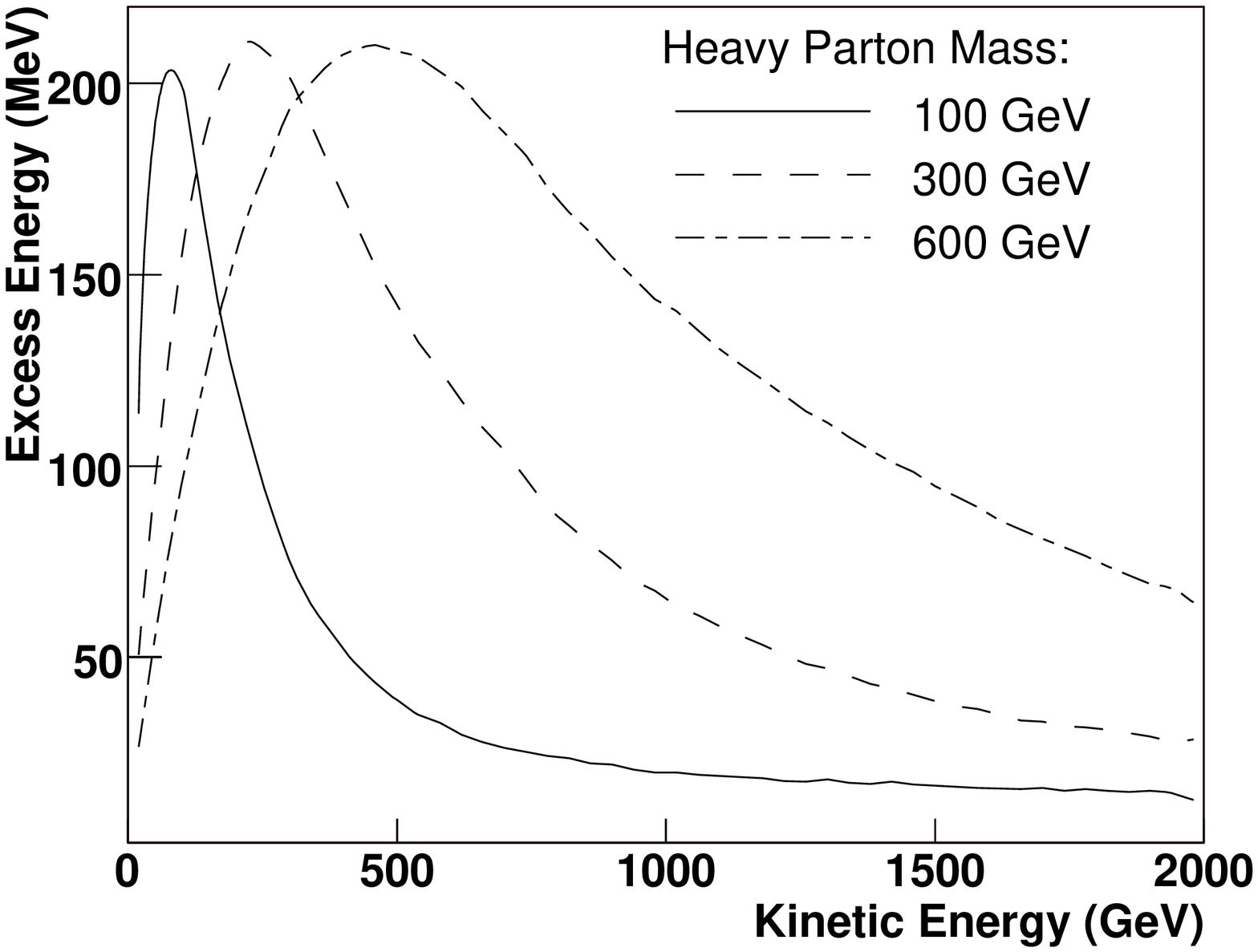,width=6cm}}
  \subfigure[Gluino hadrons in
  carbon]{\epsfig{file=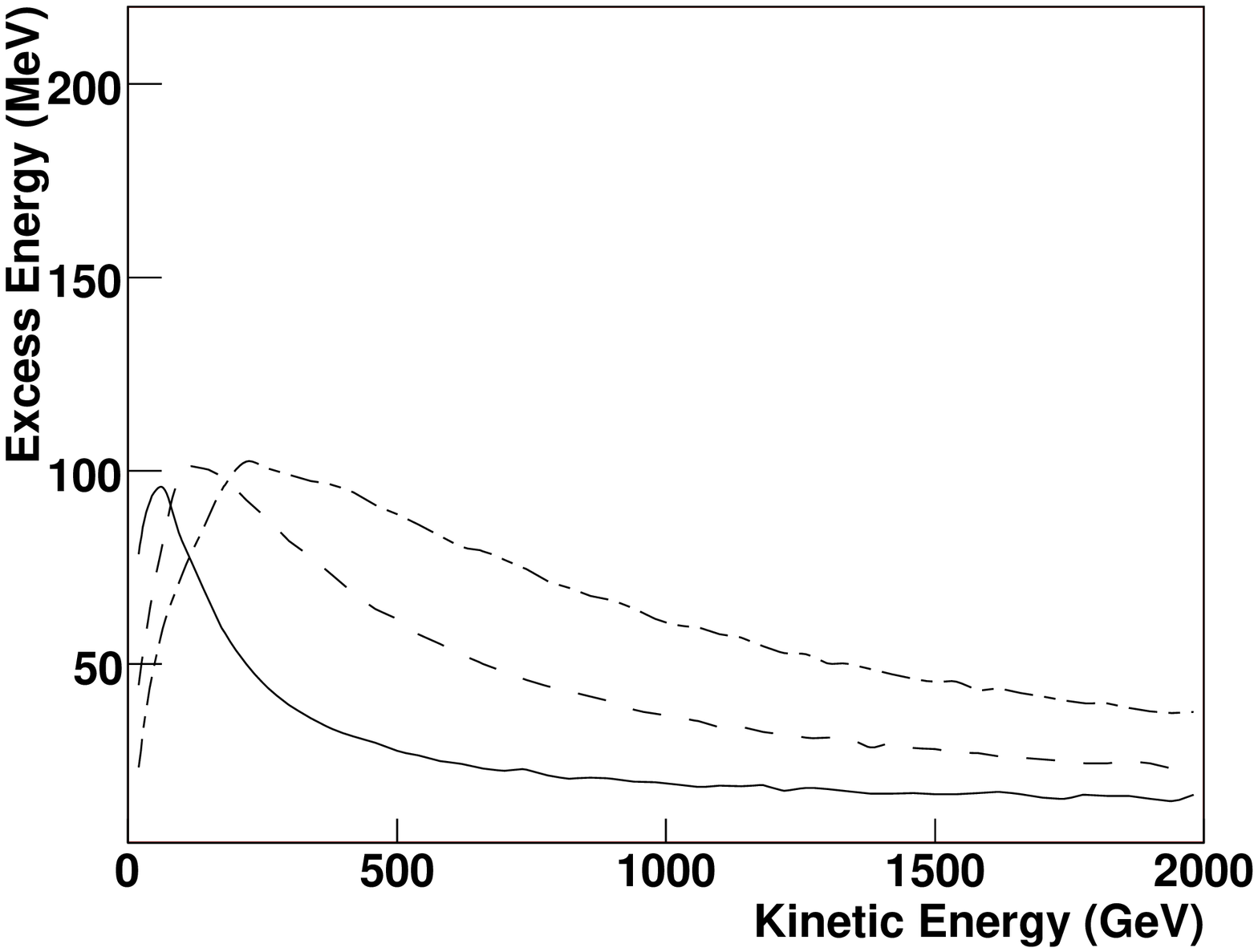,width=6cm}}
  \subfigure[Stop hadrons in
  iron]{\epsfig{file=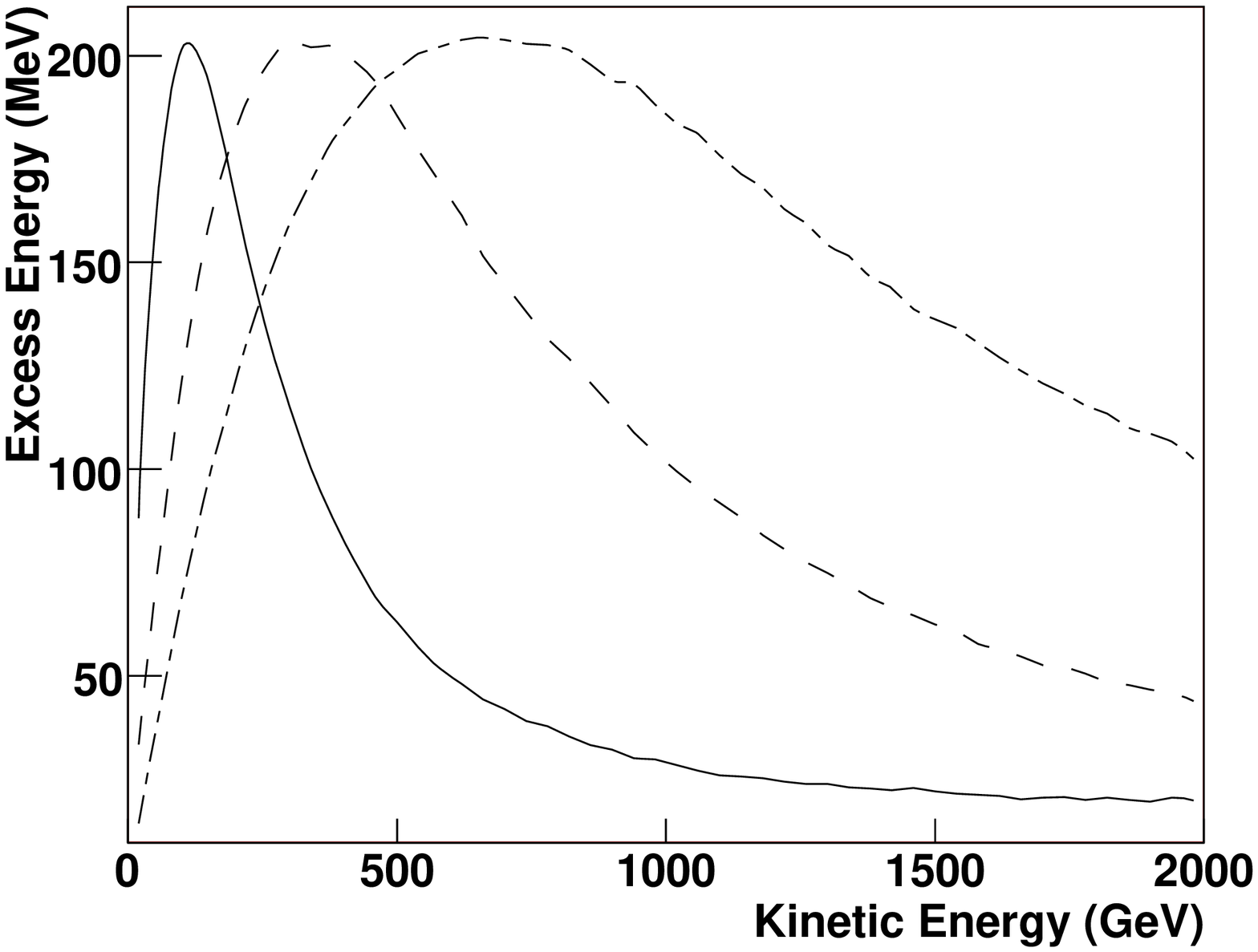,width=6cm}}
  \subfigure[Stop hadrons in
  carbon]{\epsfig{file=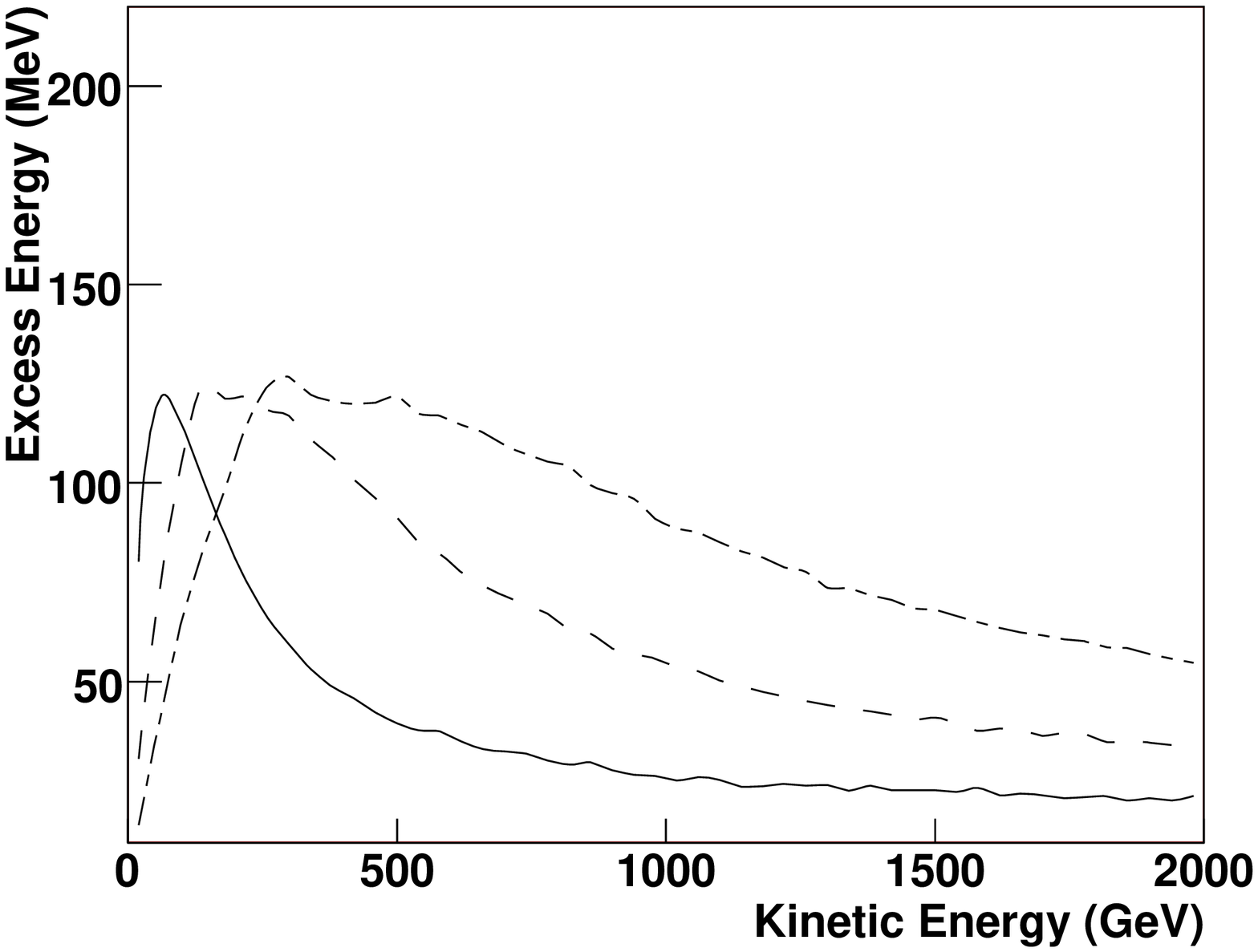,width=6cm}}
  \caption{\protect\footnotesize Local energy deposit from momentum recombination.}
  \label{fig:virt}
\end{figure}

As for the toy-model, the excess energy from the recombination is added
as a local energy deposit by Geant4. This local energy deposit is
shown in figure \ref{fig:virt} for gluino hadrons as well as stop
hadrons in iron and carbon.

It should be stressed that this is the only point at which the physics
content of the model deviates from Geant3. In Geant3 the heavy hadron
was treated as one heavy object in the kinematics treatment, and the
kinematic distributions were rescaled to take into account that not
all of the kinetic energy was available for production of secondaries.
In the present work the reaction is viewed as being between a totally
decoupled LQS and a nucleus with a subsequent re-coupling of the LQS
with the heavy parton. All differences between Geant3 and Geant4
should thus be expected to have their origin in this difference of
kinematical treatment.
 
\section{Results}
\label{sec:results}

Process lists corresponding to the cases of stable gluino hadrons and
stable stop hadrons were implemented assuming no oscillation in the
stop case. No resonance or charge exchange suppression was configured.

The first observation that can be made from simply considering the
possible processes is that a gluino meson or a stop mesonino
undergoing nuclear reactions in matter may at some point acquire a
baryon number, thus turning into a gluino baryon or a stop sbaryon
respectively. In ordinary matter it is impossible for the heavy hadron
to part with the acquired baryon number due to baryon number
conservation.  Conversely anti-stop sbaryons will tend to annihilate
with nuclear matter leading to a conversion into mesoninos.
Where not explicitly noted differently otherwise, the particles studied were
gluino baryons and stop baryons. 
In the cases where anti-stops were
explicitly studied the incident particle was chosen to be a anti-stop
mesonino.

The focus in this section will be twofold. The compatibility of the
results obtained with Geant4 to the Geant3 results will be
demonstrated, and differences between the colour octet / colour
singlet heavy parton cases will be highlighted.

\subsection{Baryonisation}
\label{sec:baryonisation}

\begin{figure}[!htbp]
  \centering
  \epsfig{file=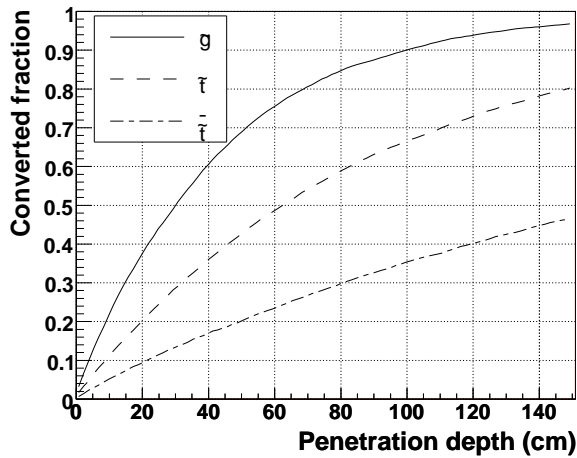,width=8cm}
  \caption{\protect\footnotesize Fraction of gluino baryons, stop
    sbaryons and anti-stop mesoninos as a function of the travelled distance in Iron. }
  \label{fig:flip}
\end{figure}
Figure \ref{fig:flip} shows the fraction of gluino mesons, stop
mesoninos and anti-stop sbaryons that have converted into gluino
baryons, stop sbaryons or anti-stop menoninos respectively as a
function of the travelled distance in iron. The heavy parton has been
given a mass of 300 \massG{} but this has little influence as long as
$M_{LQS} \ll M_{Tot}$ due to the nature of the cross section
calculation. There is no discernible difference between the toy-model
and the parametrised model, as this quantity exclusively relies on
pure phase space considerations.  Any difference would have to be
induced by nuclear binding which is a small effect at the energies
considered.  The curve representing the gluino case is seen not to be
identical to that of the previous work. In Geant3 approximately 96\%{} of the
gluino mesons had converted into baryons after travelling through 1 m
of iron \cite{Kraan:2004tz} whereas the corresponding number for
Geant4 is 90\%. There is, however a dependence of these curves on the
kinematic input distributions. The Geant4 curve was made for 300
\massG{} gluino R-hadrons generated with a flat distribution of
kinetic energies ranging from 0 to 2 TeV. Running the programme at
lower energies opens up the phase space for \pp processes relative to
that for \ppp processes. The \pp processes have a higher fraction of
baryon number changing processes than the \ppp processes leading to
earlier conversions.

Looking at the stop and anti-stop hadrons it is seen that the stop and
anti-stop hadrons take, on average, a longer distance to change their
baryon number due to the lower cross section of the smaller LQS.

\subsection{Energy Loss}
\label{sec:E-loss}

\begin{figure}[!htbp]
  \centering
  \epsfig{file=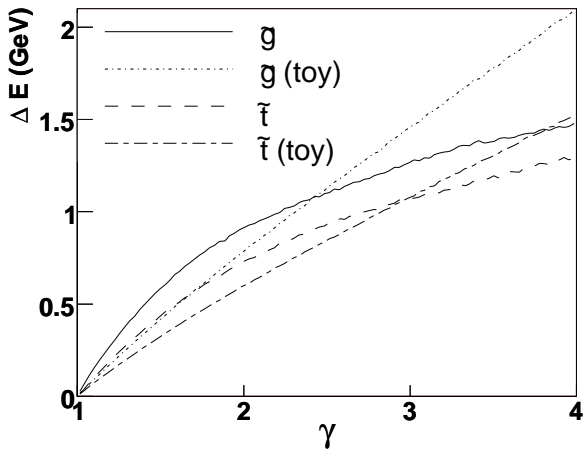,width=8cm}
  \caption{\protect\footnotesize Energy loss per hadronic interaction for gluino hadrons and stop hadrons.}
  \label{fig:eloss_gl_vs_st}
\end{figure}

Another quantity influenced by the mass of the LQS of the
stop-R-hadrons is the hadronic energy loss as can be seen in figure
\ref{fig:eloss_gl_vs_st}. The LQS in this case carries a smaller
fraction of kinetic energy and thus less energy is available in the
collision than for the gluino case. This behaviour is clear from the
plot. Considering the toy-model in comparison with the parametrised
model, one observes that the energy loss grows approximately linearly
with the $\gamma$-factor and hence with the energy. This is consistent
with the LQS carrying a constant fraction of the total energy given
the simplified treatment of the collision in this model. The
parametrised model on the other hand shows a somewhat smaller energy
loss, the rise of which decreases with energy. As the main difference
between the two models is the inclusion of the struck nucleus we can
thus here directly see the effect of the nucleus on the energy loss.

\begin{figure}[htbp]
  \centering
  \subfigure[Geant3 \cite{Kraan:2004tz}. The higher set of points is \ppp
  processes, while the lower is \pp. The central set of points is the combined one.]{\epsfig{file=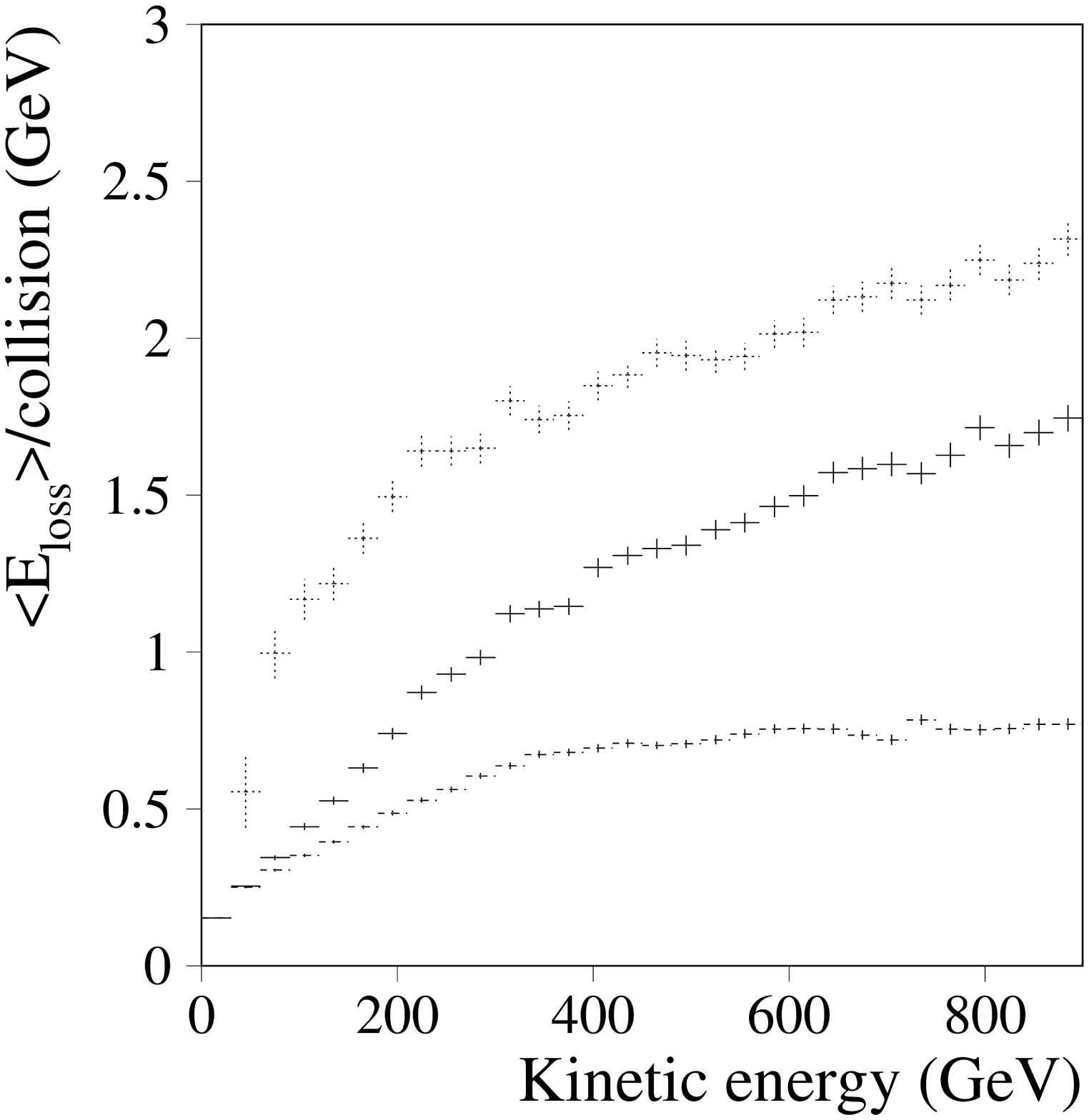,width=6cm,height=4.5cm}}
  \subfigure[Geant4 - Parametrised model]{\epsfig{file=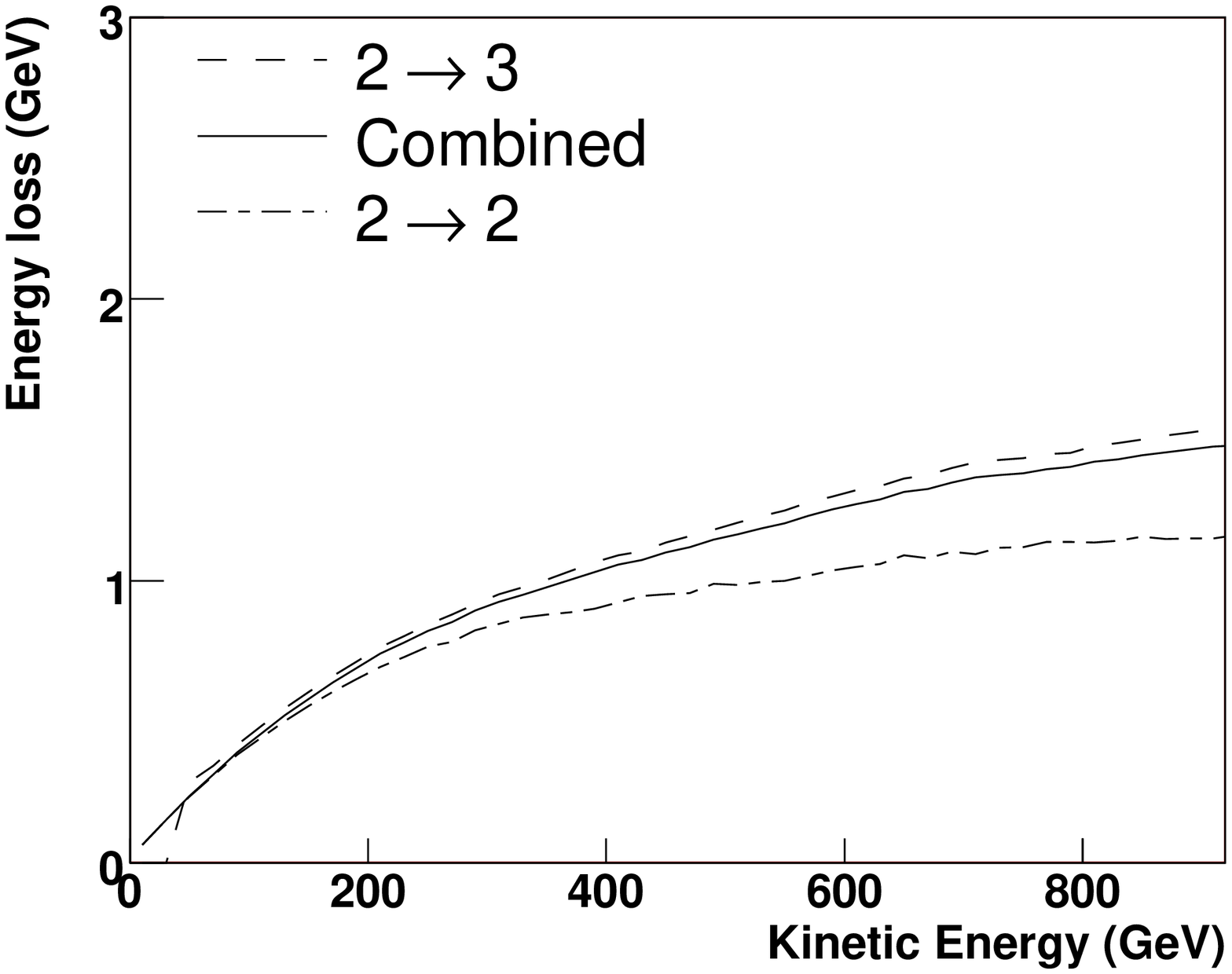,width=6cm}}
  \subfigure[Geant4 - Toy model]{\epsfig{file=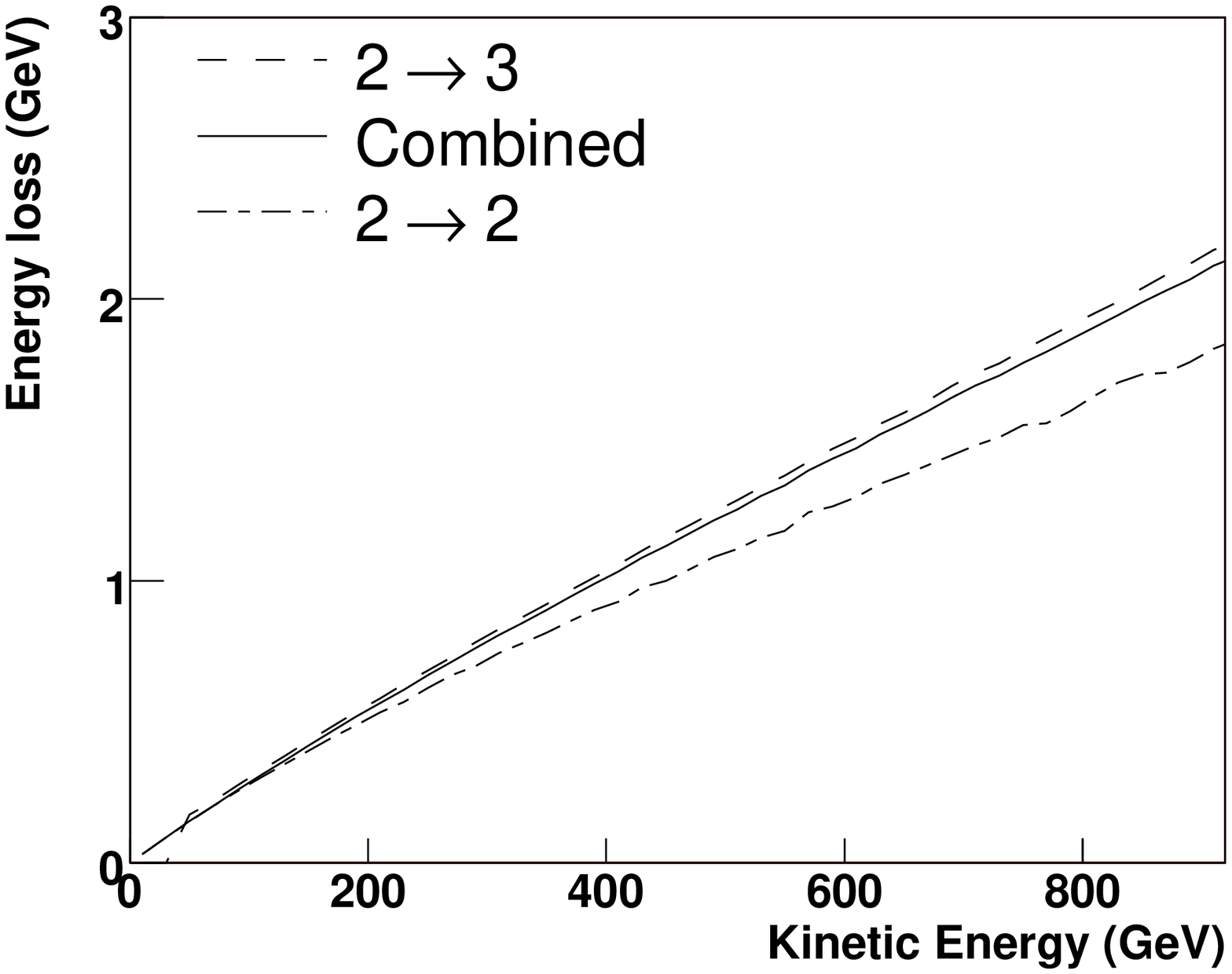,width=6cm}}
  \caption{\protect\footnotesize Energy loss per hadronic interaction for a 300 \massG{} gluino
    in iron. Both the Geant3 and Geant4 values are shown.}
  \label{fig:E_loss}
\end{figure}

Turning to figure \ref{fig:E_loss} which is a detailed comparison of
the energy loss per hadronic interaction between the models in play,
one sees immediately that the toy-model is not consistent with the
previous work in either a quantative or qualitative sense, as is
expected. It \emph{does}, however, give numbers within the same order
of magnitude. We thus see clearly for both distributions the changes
in phenomenology induced by the treatment of the struck nucleus and
the consequent changes in kinematics.
Looking at the parametrised model the energy loss is closer, but not
identical, to that predicted with Geant3 model. The difference is
expected because the kinematics tratment because the kinematics
treatment is somewhat different between the Geant3 and Geant4 versions
of this model as discussed in section \ref{sec:parmmodel}.
 
\begin{figure}[!htbp]
  \centering
  \subfigure[Ionisation loss\label{fig:IonE_loss}]{\epsfig{file=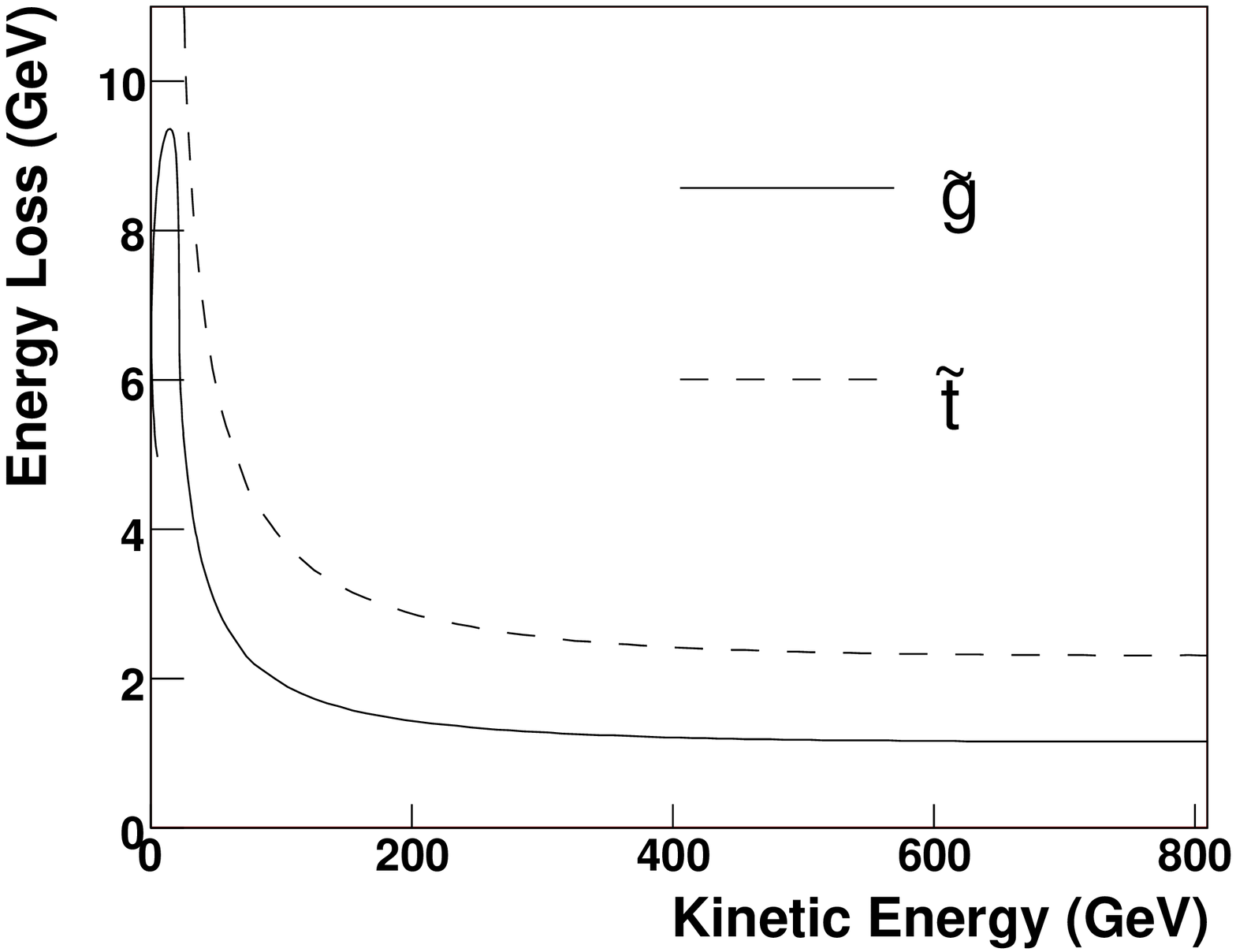,width=5.5cm}}
  \subfigure[Hadronic energy loss]{\epsfig{file=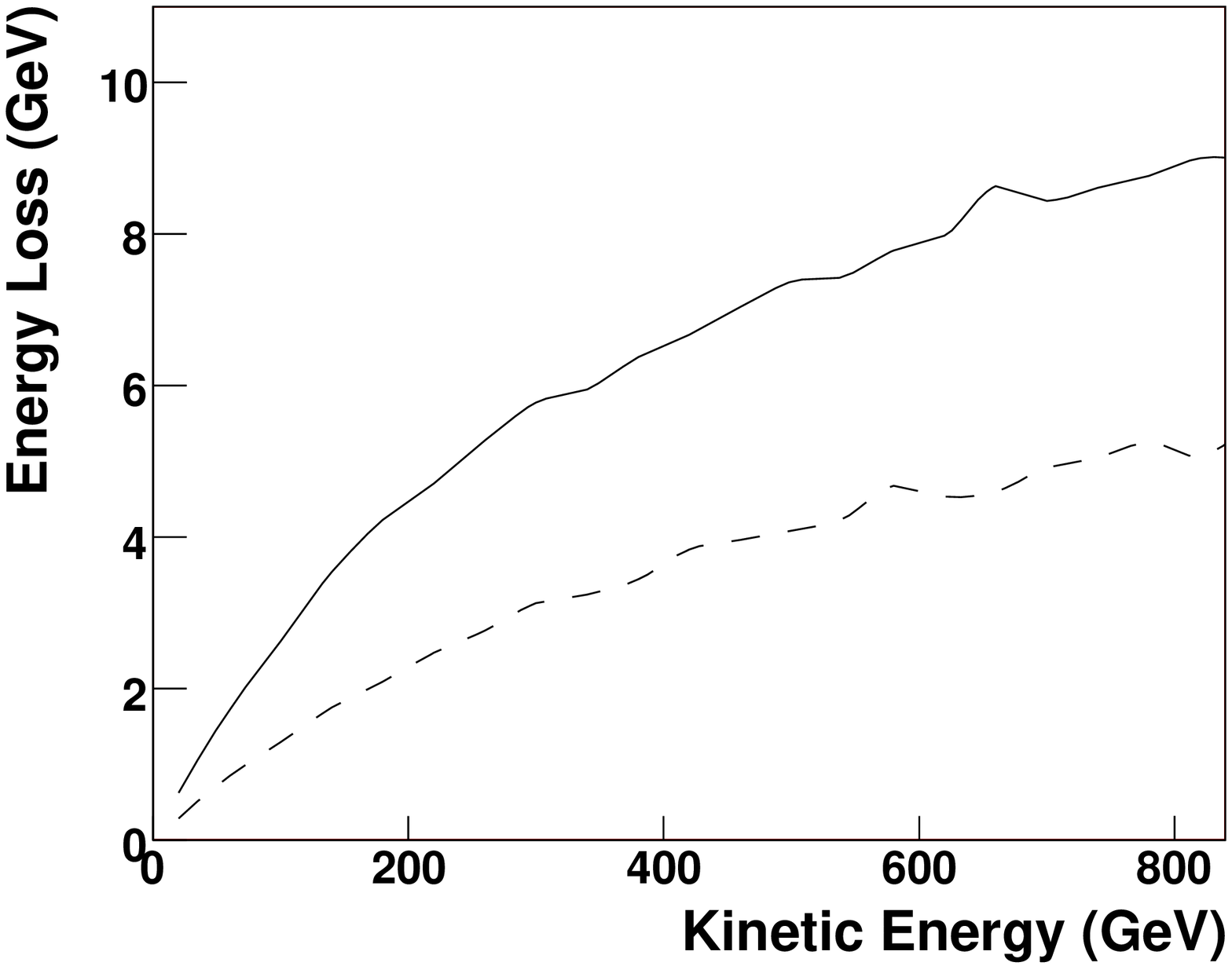,width=5.5cm}}
  \subfigure[Total energy loss]{\epsfig{file=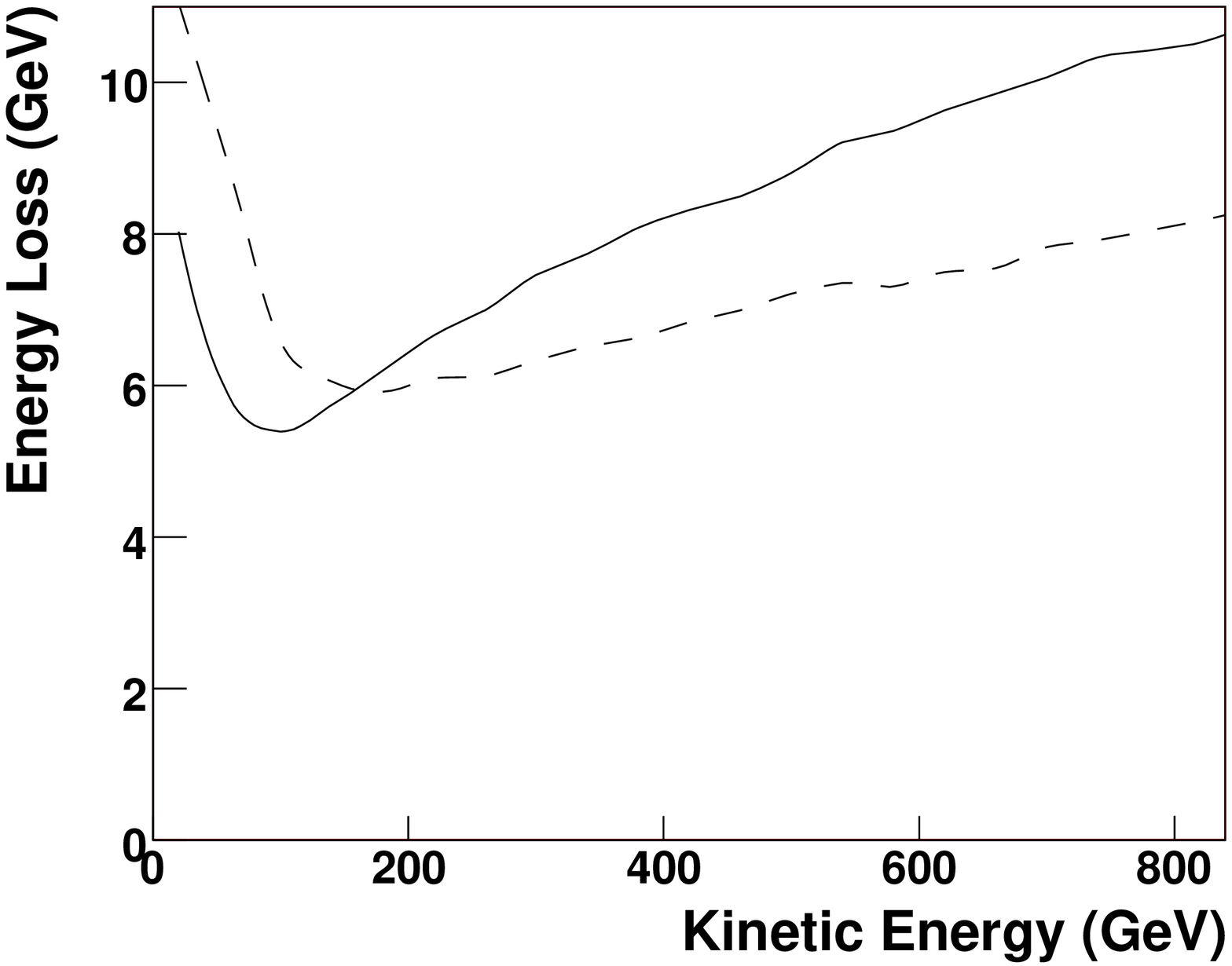,width=5.5cm}}
  \caption{\protect\footnotesize Absolute energy loss in 1 m of iron
    for a 300 \massG{} gluino / stop hadron. In the Geant4 plots the
    numbers corresponding to gluino hadrons are represented by the
    fully drawn line whereas the stop hadrons are represented by a
    dotted line. }
  \label{fig:abs_E_lossG4}
\end{figure}

The energy loss is seen to be somewhat higher for the \pp processes in
Geant4 relative to what is predicted in Geant3 while the opposite is
true for \ppp processes. The two models are still roughly compatible,
however, and nothing here suggests a drastic difference in
phenomenology between the two models.
To corroborate this observation the energy loss in 1 m of iron for a
300 \massG{} heavy parton is shown in figure \ref{fig:abs_E_lossG4}. The
values were calculated using the parametrised model only.
\begin{figure}[!htbp]
  \centering
  \epsfig{file=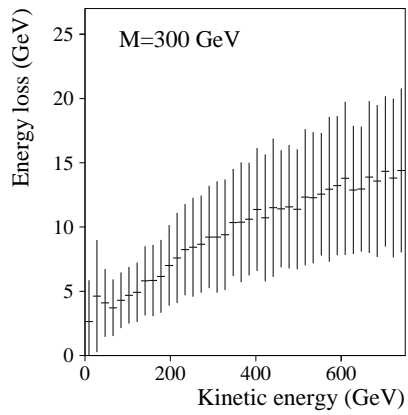,width=6cm}
  \caption{\protect\footnotesize Geant3 hadronic energy loss in 1m of
    iron for a 300 \massG{} gluino hadron.\cite{Kraan:2004tz}}
  \label{fig:G3E_loss}
\end{figure}

In the figure one can see both the ionisation and hadronic energy
losses, and the combination for a 300 \massG{} heavy parton. This is
in comparison with the values from Geant3 which are shown in figure
\ref{fig:G3E_loss}. The values are seen to be compatible. As the
average kinetic energy carried by the LQS is smaller for the stop case
than for the gluino case as previously mentioned, the hadronic energy
loss is correspondingly smaller and the contribution from ionisation
is thus clearly more visible for the stop case than for the gluino
case. The effect is amplified by the fact that by simple quark
counting a stop sbaryon will more often have double charge than what
is the case for gluino baryons. The charge enters the Bethe Bloch
formula squared, leading to a substantially higher energy loss due to
ionisation. This is indeed observed in figure \ref{fig:IonE_loss}.

We can thus conclude that the physics content of the Geant3 and the
Geant4 models is the same, with one quantitative difference induced by
a conceptual difference in the handling of the kinematics.


\subsection{dE/dx}

Turning to a classic observable, energy loss distributions were
calculated as a function of distance crossed by the R-hadrons in iron.
The simulations were produced in the case of gluino baryons, stop
sbaryons and anti-stop mesoninos to simplify matters in relation to
section \ref{sec:baryonisation}.
\begin{figure}[!htbp]
  \centering
  \epsfig{file=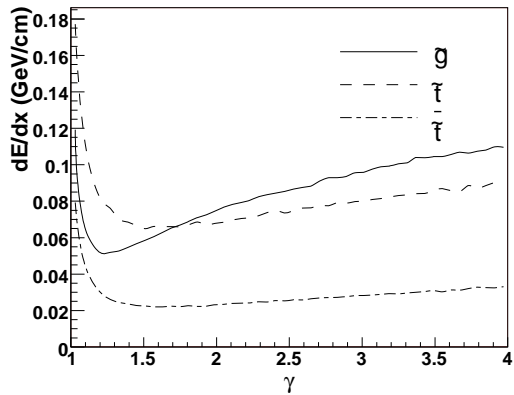,width=7cm}
  \caption{\protect\footnotesize dE/dx distributions for a 600
    \massG{} sparticle mass comparing the stop and anti-stop to the
    gluino case.}
  \label{fig:dedx}
\end{figure}

As can be seen from figure \ref{fig:dedx} there is a difference in the
shape of the distribution between the gluino and the stop/anti-stop
case while the main difference between the stops and the anti-stops is
the overall energy deposition. The difference in shape stems from the
cross-over where the hadronic energy loss starts to dominate over the
EM energy loss.
\begin{figure}[!htbp]
  \centering
  \subfigure[Stops]{\epsfig{file=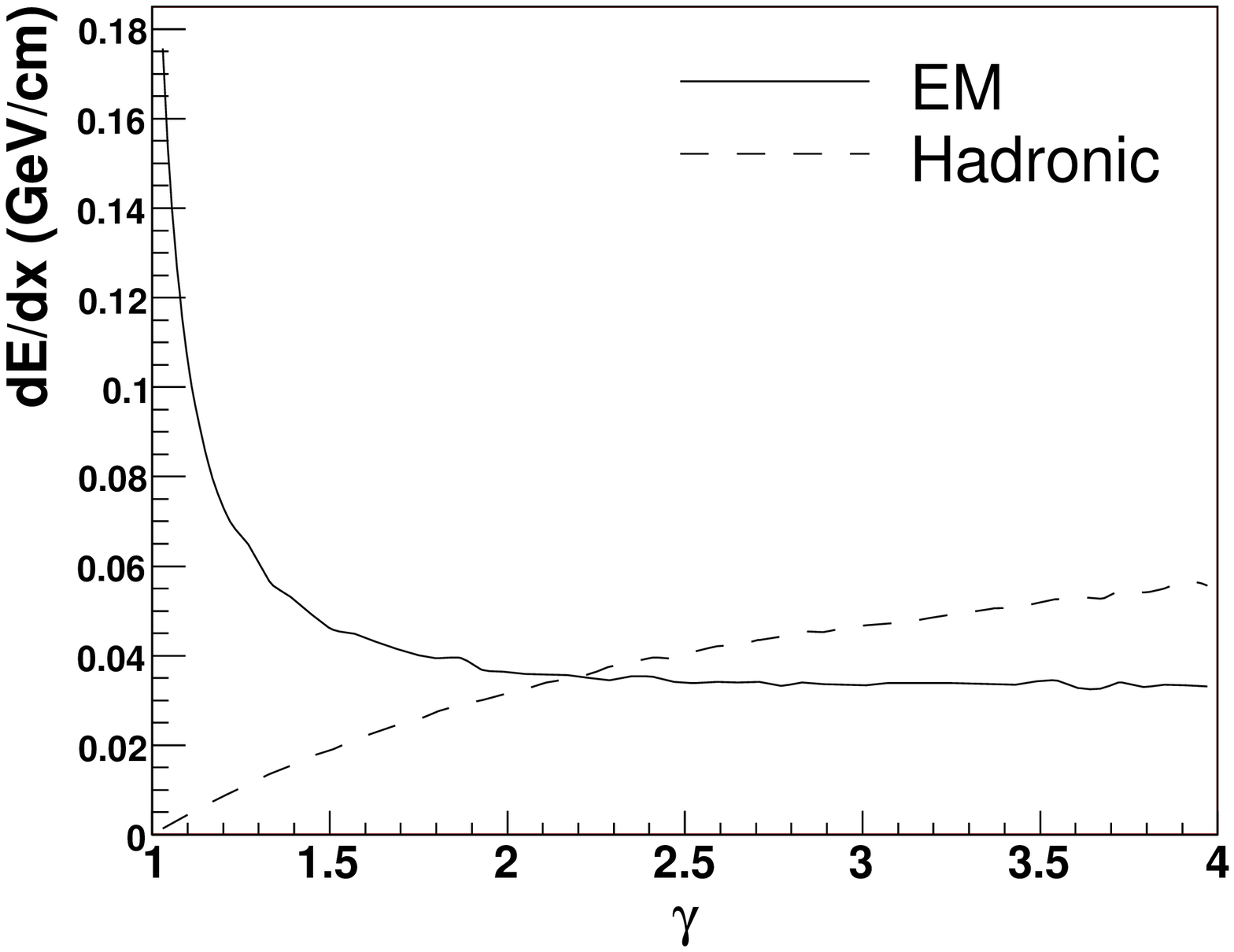,width=6cm}}
  \hfill
  \subfigure[Anti-stops]{\epsfig{file=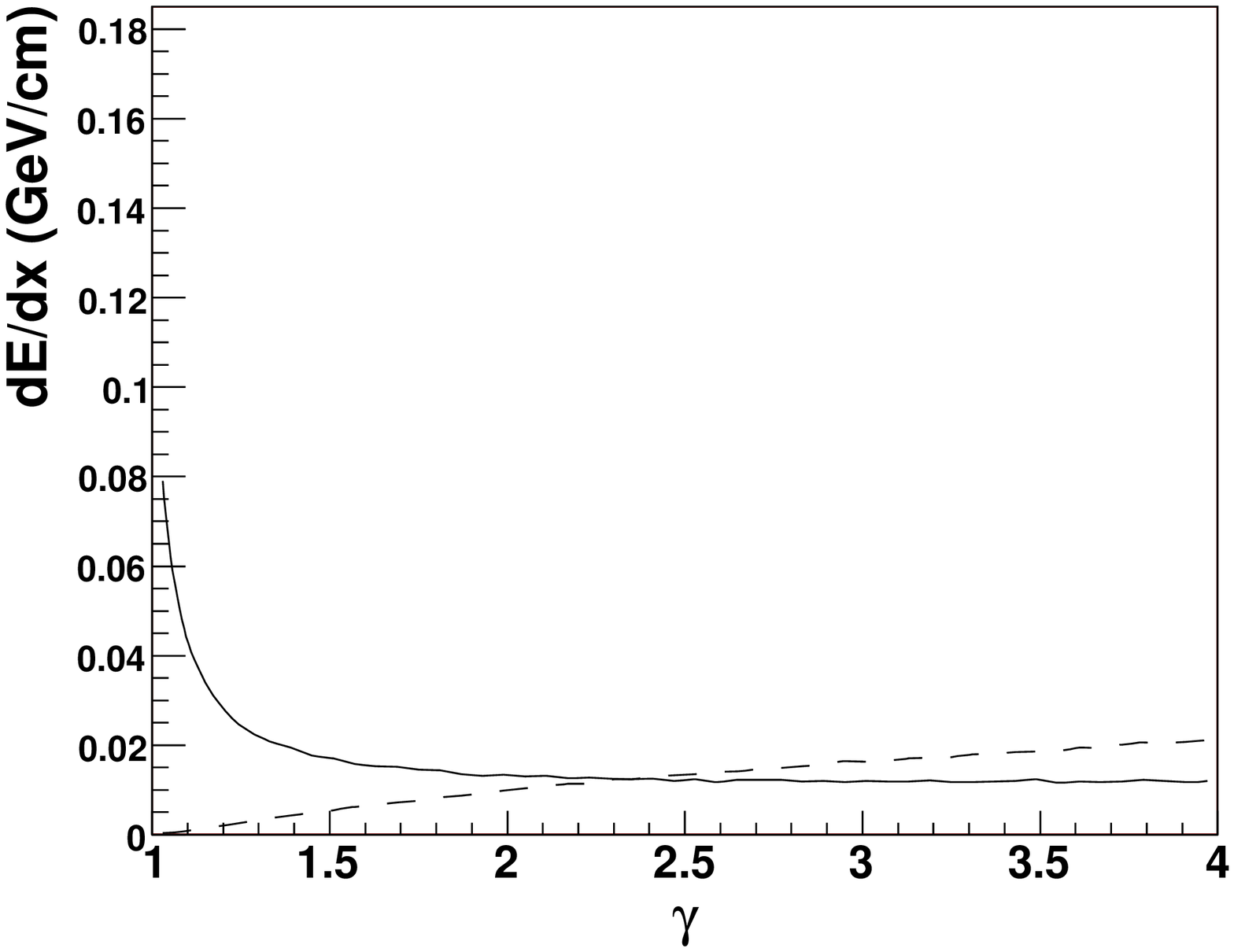,width=6cm}}  
  \hfill
  \subfigure[Gluinos]{\epsfig{file=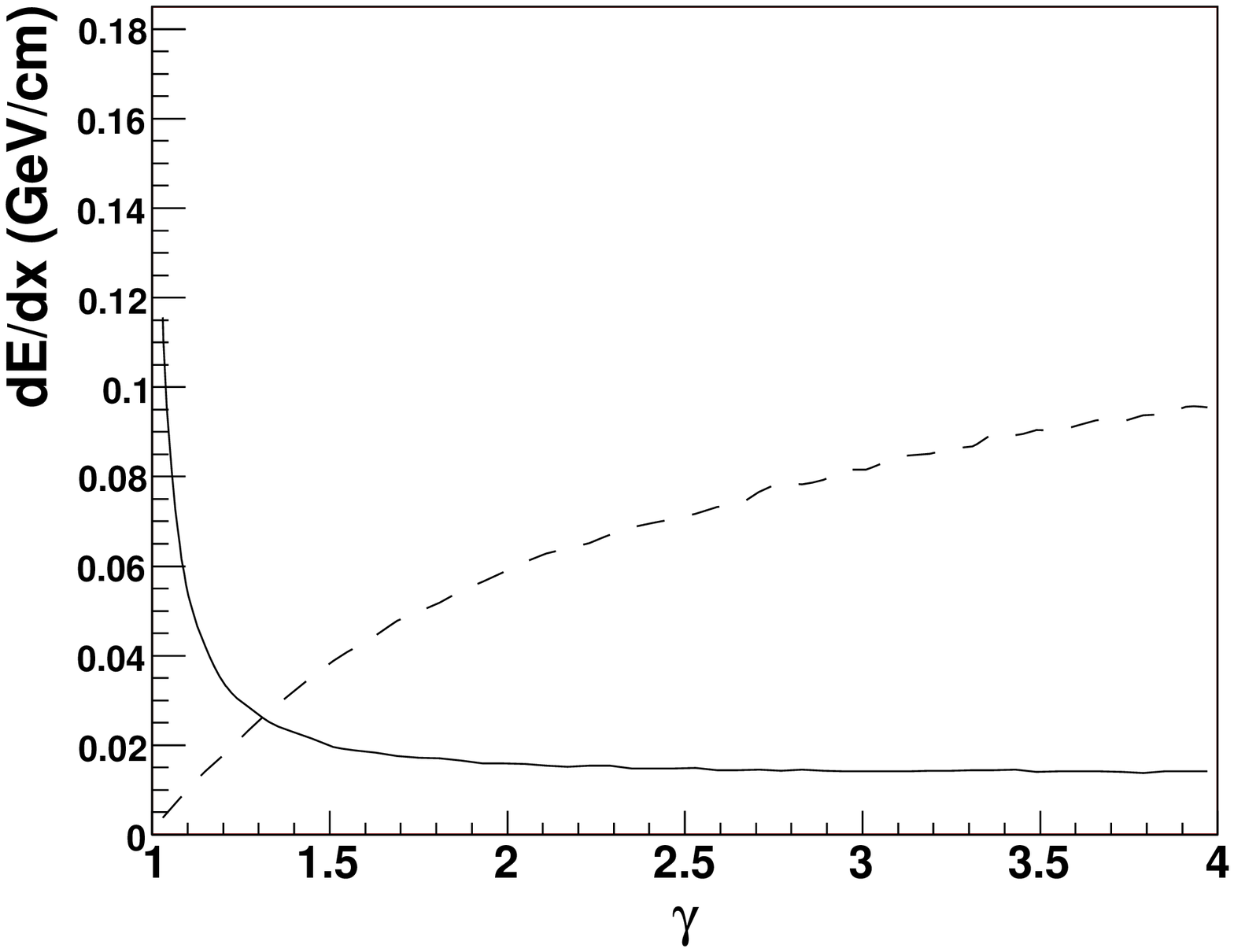,width=6cm}}  
  \caption{\protect\footnotesize Cross-over between EM and hadronic energy loss for stops and gluinos}  \label{fig:xover}
\end{figure}

Figure \ref{fig:xover} shows dE/dx for all three kinds of
particles where the energy loss has been split into an EM contribution
and a hadronic contribution. It is clear from the figure that stops
and anti-stops are very similar apart from the overall size of the
energy loss. Furthermore the onset of the hadronic energy loss is a
lot sharper in the gluino case due to the larger cross section and
energy loss per interaction as already discussed in section
\ref{sec:E-loss}.

As R-hadrons are often produced in pairs, the difference in
normalisation of the dE/dx distributions for squarks and anti-squarks
could prove to be a powerful estimator to distinguish between a gluino
and a squark signal. If dividing a sample of R-hadron candidates into
a positively and negatively charged sample resulted in large
differences in energy depositions, one would have a powerful argument
for a charged colour triplet hypothesis. Conversely, if no difference
was found or if the samples proved to be ill defined, one would be
able to reject the squark R-hadron scenario for at least the case of
no oscillations.


%
%



\subsection{Shower shapes}
\label{sec:shower}

\begin{figure}[!htbp]
  \centering
  \epsfig{file=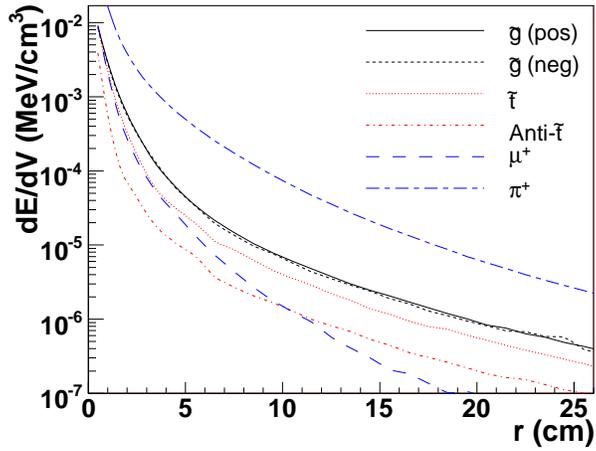,width=8cm} 
  \caption{\protect\footnotesize Transverse shower profile for 300
    \massG{} R-hadrons compared with muons and pions.}
  \label{fig:show}
\end{figure}

The electromagnetic and hadronic interactions of R-hadrons and their
collision reaction products will give rise to showers of secondary and
tertiary particles. The profiles of these showers might provide a
handle on finding and identifying R-hadrons. Samples were generated in
which beams of particles were shot into 10 m of iron. The particles
were generated with a flat distribution in kinetic energy up to 1 TeV.
Figure \ref{fig:show} shows the energy deposition density transverse
to the incident direction of the particles. Hadronic interactions of
secondaries were treated using the a standard Geant4 physics list
(QGSP).
\begin{figure}[htbp]
  \centering
  \epsfig{file=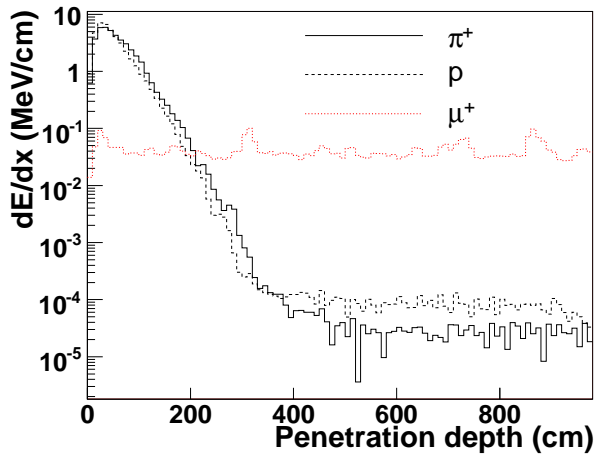,width=8cm}
  \caption{\protect\footnotesize Longitudinal shower profile of pions
    and protons compared to muons.}
  \label{fig:longshow_pi_p}
\end{figure}

Figure \ref{fig:longshow_pi_p} sets the scale of the energy
depositions. It is seen that hadrons will mostly deposit all their
energy within the first 2-3 m at a scale that is two orders of
magnitude beyond typical values for muons.
\begin{figure}[htbp]
  \centering
  \epsfig{file=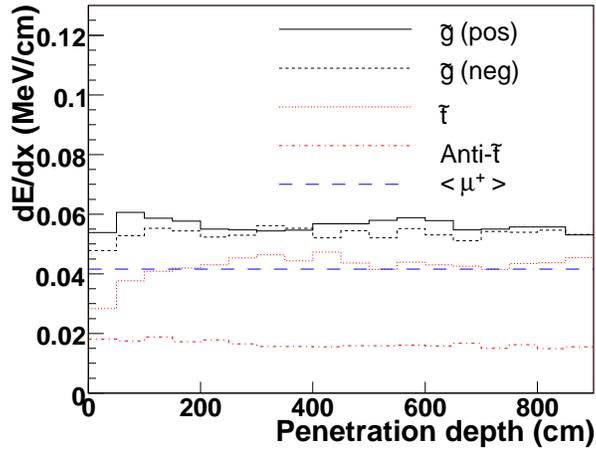,width=8cm}
  \caption{\protect\footnotesize Longitudinal shower profile of muons and 300 \massG{} R-hadrons generated with the same flat kinetic energy distribution}
  \label{fig:longshowR}
\end{figure}

In figure \ref{fig:longshowR} the various types of R-hadrons are
compared to the \emph{average} muon energy deposition. It is seen that
the level of energy deposition is comparable between the muons and the
R-hadrons.
A detailed comparison between the different species of R-hadrons shows
little difference between gluino R-baryons generated with positive or
negative charge. This is to be expected as after a few hadronic
interactions the R-hadron will have no memory of its initial charge.
Also the sign of the charge plays no role neither in the Bethe-Bloch
formula or the hadronic interactions.  As expected the lower cross
section of the anti-stop states lead to a smaller overall energy
deposition than for any of the other cases.

Turning to the transverse shower profiles, figure \ref{fig:show} shows
the energy deposition density at a distance $r$ from the incident
direction of the R-hadron. R-hadrons are compared to muons and pions
in order to set a natural scale. From the figure we see again the same
differences between the different types of R-hadrons as was the case
for the longitudinal profiles. The pions deposit substantially more
energy than R-hadrons.  This is due to the fact that they deposit all
their energy, while the R-hadrons punch through as seen in figures
\ref{fig:longshow_pi_p} and \ref{fig:longshowR}. The muons tend to
produce a narrower shower than the R-hadrons. This is consistent with
the muons not having hadronic interactions. It thus appears that
R-hadron showers deposit substantially less energy than ordinary
hadrons and that the energy deposition is very much alike to that of
muons apart from the fact that the shower is broader. We may add to
this observation the fact that an R-hadron traversing a calorimeter
may at times be neutral, leading to gaps in the energy deposition when
looking at the longitudinal profile of the shower.



\section{Conclusions}
\label{sec:concl}

A method has been presented to simulate the phenomenology of heavy
coloured long-lived particles that show up in many novel extensions to
the SM. Also with the stop and gluino study we have demonstrated for
the first time how predictions can be made to discern between colour
triplet and octet states from their interactions with matter. Many
observations from this paper may be transferred directly to other
models. More novel constructs carrying other combinations of SM
charges than quarks and gluons will require careful consideration as
to which processes will occur in the interaction with matter. In
general, in order to use this model in other scenarios, it is needed
to check whether or not the considerations presented here apply.

\section{Acknowledgements}
\label{sec:ack}

The authors wish to acknowledge Dennis Wright (SLAC) for his
assistance in weeding out the flavour dependence in the parametrised
model of Geant4. Also we wish to thank Troels Petersen for supplying
the Dalitz plot class used in the toy model of section
\ref{sec:toymodel}. Finally we wish to thank Torbj\"orn Sj\"ostrand
for providing Pythia code to handle gluino and stop hadronisation as
well as a list of gluino and stop hadrons with masses and PDG
compatible codes.

\end{document}